\documentclass[aps,pra,epsfigure,twocolumn]{revtex4}

\usepackage{dcolumn}
\usepackage{bm}
\usepackage{graphicx}
\usepackage{amsmath}
\usepackage{latexsym}
\usepackage{amsfonts}
\usepackage{amssymb}
\usepackage{array}
\usepackage{epsfig}

\newcommand{\ket}[1]{\left\vert#1\right\rangle}

\newcommand{\one}{\mbox{$1 \hspace{-1.0mm}  {\bf l}$}}
\newcommand{\pro}[2]{\left\vert#1\right\rangle\left\langle#2\right\vert}

\newcommand{\bra}[1]{\left\langle#1\right\vert}

\begin{document}


\title{Cross-Kerr-based information transfer processes}

\author{Jinhyoung Lee$^1$, M. Paternostro$^2$, C. Ogden$^2$, Y. W. Cheong$^1$, S. Bose$^3$ and M. S. Kim$^2$}
\affiliation{$^1$Department of Physics, Hanyang University, Sungdong-Gu,
  133-791 Seoul \\ and Quantum Photonic Science Research Center, Hanyang  University, Seoul 133-791, Korea\\
 $^2$School of Mathematics and Physics, Queen's University, Belfast BT7
 1NN, United Kingdom\\
  $^3$School of Physics and Astronomy, Gower Street, London WC1E 6BT,
  United Kingdom}

\date{\today}

\begin{abstract}

The realization of nonclassical states is an important task for many applications of quantum information processing. Usually, properly tailored interactions, different from goal to goal, are considered in order to accomplish specific tasks within the general framework of quantum state engineering. In this paper we remark on the flexibility of a cross-Kerr nonlinear coupling in hybrid systems as an important ingredient in the engineering of nonclassical states. The general scenario we consider is the implementation of high cross-Kerr nonlinearity in cavity-quantum electrodynamics. In this context, we discuss the possibility of performing entanglement transfer and swapping between a qubit and a continuous-variable state. The recently introduced concept of entanglement reciprocation is also considered and shown to be possible with our scheme. We reinterpret some of our results in terms of applications of a generalized Ising interaction to systems of different nature.

\end{abstract}


\maketitle


\section{Introduction}

Quantum state engineering is a necessary preliminary step in many applications of quantum information processing (QIP). The realization of nonclassical states of discrete as well as continuous variables (CV) is often required as an off-line resource in protocols for quantum computation and communication. As an example, the preparation of Schr\"odinger cat states of a CV system is one of the key ingredients in the performance of coherent state quantum computation~\cite{vanenk,jacob}. Very recently, efficient proposals for the realization of Schr\"odinger cat states have been put forward~\cite{gatti}. On the other hand, an entangled coherent state of two field modes~\cite{sanders} serves as a quantum channel in the realization of teleportation-based coherent state-quantum computation~\cite{ralph,ioMyungKnight}. In these last years, the use of hybrid quantum registers made up of systems having different nature ({\it i.e.} spanning Hilbert spaces of different dimensions) has been recognized to be a useful way to improve the flexibility of a physical device. Both the manipulability and the robustness against decoherence in computation and communication are often improved by a hybrid setup. Therefore, the necessity of performing quantum state engineering involving systems of different nature is raising increasing interest in the QIP community.

A usual requirement in the context of quantum state engineering of coherent states is the availability of efficient sources of nonlinear interactions. Unfortunately, the current state of the art is such that only small nonlinear rates can be attained. This represents a big hindrance not only for the creation of nonclassical states but also for single-photon quantum computation. Ways to nearly-deterministically bypass this bottleneck have been envisaged with the simulation of nonlinear interaction by photodetection~\cite{KLM}. More recently, Munro and coworkers~\cite{munro} showed an efficient way to perform all-optical quantum computation using single photons by adopting the quantum non-demolition (QND) measurement technique using a low cross-Kerr nonlinearity and strong amplitude coherent field.

Even though this represents a major step forward in the field of manipulating the state of travelling-light waves, the proposal in~\cite{munro} is not the only situation where the use of nonlinearity-based QND can be successfully exploited. In cavity-quantum electrodynamics (cavity-QED), a very similar technique involving a bosonic system in a coherent state and a fermionic system in a two-dimensional state has already been experimentally demonstrated. In particular, Bertet {\em et al.}~\cite{bertet} measured the Wigner distribution of a Fock state based on the theoretical proposals~\cite{englert}, which exploit cross-Kerr-based QND techniques and, very recently, Maioli {\em et al.} performed Rydberg atom counting based on QND measurement~\cite{maioli}. It is worth stressing that, differently from a running field, the nonlinearity rate achievable in cavity-QED is large enough not to require exceedingly long interaction times or intense light fields.

In this paper, we investigate further on the use of cross-Kerr QND interactions by suggesting theoretical protocols for the implementation of effective quantum state engineering in a cavity-QED setup. In all the schemes suggested here, the synergy between two-level and infinite dimensional systems is the key element. We address explicitly the transfer of entanglement between subsystems of different nature, thoroughly analyzing the conditions for the maximal transfer of entanglement. This results in the proposal for entanglement transfer from discrete to CV systems and the adaptation of the recently proposed protocol for {\it entanglement reciprocation}~\cite{noireciprocation}, which achieves a complete ebit of a two-qubit state starting from a CV entangled state. The inherently hybrid nature of the cross-Kerr interactions treated here allows us to design a simple scheme for entanglement swapping~\cite{swapping} between discrete and CV entangled states. The possibility of using the resulting state for building up effective quantum repeaters~\cite{repeaters} for long-range communication is briefly sketched.

The remainder of this paper is organized as follows. In Section~\ref{interazione} we briefly describe the formal approach to the interaction between an atom and a coherent state of a cavity field-mode at the basis of the effective cross-Kerr interaction. Section~\ref{state} is devoted to information transfer, which serves as a basic building block to be used for the other tasks analized in our work. These are mainly entanglement transfer, approached in Section~\ref{transfer} and entanglement reciprocation, which is the center of Section~\ref{reciprocation}. Section~\ref{multiple} shows that the trasnfer and reciprocation scheme can be generalized in such a way that an initial entanglement corresponding to $n$ ebits can be mutually exchanged between matter and light. Finally, in Section~\ref{swapping} we fully exploit the hybrid nature of the QND model used here to suggest a suitably modified version of an entanglement swapping protocol. Section~\ref{conlcusioni} remarks on the ingredients required in the physical implementation of our protocols and outlines our main results.


\section{Interaction model}

\label{interazione}

In this Section, we briefly describe how the cross-Kerr effect is achieved in a cavity-QED system. We consider a two-level atom with excited and ground states $|e\rangle$ and $|g\rangle$ respectively, interacting with a single-mode cavity field with bosonic annihilation (creation) operators $\hat{a}$ ($\hat{a}^\dag$). The atomic transition frequency is $\omega_0$ and the frequency of the cavity field mode is $\omega$ while the interaction rate between the two subsystems is indicated by $\lambda$. In the interaction picture and under the rotating-wave and electric dipole approximations, the Hamiltonian of the system is

\begin{equation}
\hat{H}=\frac{1}{2}\delta(|e\rangle\langle e|-|g\rangle\langle g|)+\lambda(\hat{a}^\dag|g\rangle\langle e| +\hat{a}|e\rangle\langle g|),
\label{hamiltonian}
\end{equation}

where $\delta=\omega_0-\omega$ denotes the detuning between the two-level atom and the cavity field. In the ordered atomic basis
$\langle e|=(1,0)$, $\langle g|=(0,1)$, the time-evolution operator $\hat{U}=\mbox{e}^{-i\hat{H}t}=\left(\begin{array}{cc}
\hat{U}_{11} & \hat{U}_{12} \cr
\hat{U}_{21} & \hat{U}_{22} \cr
\end{array}\right)$ is given by the following explicit expression~\cite{stenholm}
\begin{equation}
\label{evol-op}
\hat{U}=
\begin{pmatrix}
\cos \hat\Omega_{n+1}t -i{\delta}\frac{\sin\hat\Omega_{n+1}t}{2\hat\Omega_{n+1}}&-i\lambda\hat{a}\frac{\sin\hat\Omega_nt}{\hat\Omega_n}\\
-i\lambda\hat{a}^\dag\frac{\sin\hat\Omega_{n+1}t}{\hat\Omega_{n+1}}&\cos \hat\Omega_n t + i{\delta}\frac{\sin\hat\Omega_n t}{2\hat\Omega_n}
\end{pmatrix}.
\end{equation}


Here, we have introduced the effective Rabi frequency
\begin{equation}
\hat\Omega_n=\sqrt{({\delta^2}/{4})+\lambda^2\hat{n}}
\label{rabi}
\end{equation}
with the photon-number operator $\hat{n}=\hat{a}^\dag\hat{a}$. Inspired by the idea of QND interactions~\cite{braginsky}, Brune {\em et al.}~\cite{brune} proposed an experimental realization in cavity-QED using the model whose evolution is described by Eq.~(\ref{evol-op}). For this purpose, a dispersive interaction characterized by a large detuning $\delta$ has been considered rather than a resonant one. These are exactly the sort of working conditions we want to adopt in this paper. Under the condition ${\delta^2}/{4}\gg\lambda^2\bar{n}$, $\lambda^2\Delta n$, where $\bar n$ and $\Delta n$ are respectively the average photon number of the cavity field and the corresponding variance, the Rabi frequency is dominated by the detuning and only the diagonal terms in Eq.~(\ref{evol-op}) are non-negligible, reading $$\hat{U}_{11}\approx\mbox{e}^{-i\hat{\Omega}_{n+1}t},\hskip1.0cm\hat{U}_{22}\approx\mbox{e}^{i\hat{\Omega}_nt}.$$ In order to better gather the explicit form of the effective time-evolution operator, we consider the approximation $\hat\Omega_n\approx \frac{\delta}{2}+\frac{\lambda^2}{\delta}\hat{n}$ and move to a frame rotating at a frequency equal to the detuning $\delta$. In these conditions, the evolution operator Eq.~(\ref{evol-op}) of the atom-field system becomes
\begin{equation}
\hat{U}_r\equiv\mbox{e}^{-i\frac{\lambda^2}{\delta}t (\hat{n}+1)}|e\rangle\langle e|+\mbox{e}^{i\frac{\lambda^2}{\delta}t \hat{n}}|g\rangle\langle g|
\label{evol-op-2}
\end{equation}
which clearly shows that no energy exchange between the atom and the field occurs. Just the phase of the atomic state changes depending on the intensity of the cavity field mode. This is the key for the QND interaction not to destroy the photon number during its measurement. This is in perfect analogy with the evolution operator for the cross-Kerr coupling $\hat{U}_{ck}=\mbox{e}^{i\kappa t\hat{a}^\dag\hat{a}\hat{b}^\dag\hat{b}}$ between two modes $a$ and $b$ with the coupling strength $\kappa$.  A big technical difference between these two models is that, while $\kappa t$ is extremely small in current optical-fibers (frequently used in order to achieve photon-photon coupling, in running-wave configurations), ${\lambda^2 t}/{\delta}$ can be large large enough to guarantee an effectively large phase-shift within the decoherence time of the system at hand~\cite{brune}. This is an important reason which makes the cavity-QED realization of QND attractive.  Recently, based on this dispersive interaction between a field and an atom, Wang and Duan~\cite{wang} proposed to engineer Schor\"odinger cat states and Toscano {\em et al.}~\cite{toscano} suggested a Heisenberg-limited measurement of displacements and rotations.

Throughout the paper, we assume that the cavity is initially prepared with a coherent state $|\alpha\rangle$ of amplitude $\alpha\in\mathbb{C}$ which is routinely done, experimentally, by coupling an external coherent field with the cavity.  As a coherent state is a superposition of number states with the Poissonian weight, $|\alpha\rangle=\mbox{e}^{-\frac{|\alpha|^2}{2}}\sum_{n=0}^\infty\frac{\alpha^n}{\sqrt{n!}}|n\rangle$, applying the evolution operator in Eq.~(\ref{evol-op-2}) to the joint state of discrete and CV systems for the interaction time ${\lambda^2 t}/{\delta}={\pi}/{2}$, we find that
\begin{equation}
|e\rangle|\alpha\rangle\stackrel{\hat{U}}{\longrightarrow} -i|e\rangle|-i\alpha\rangle,~~
|g\rangle|\alpha\rangle\stackrel{\hat{U}}{\longrightarrow} |g\rangle|i\alpha\rangle,
\label{transform}
\end{equation}
which shows the conditional phase-shift achieved by the initial coherent state. These transformations will be frequently used throughout this paper as the central ingredients of our protocols.


\section{Information transfer}

\label{state}

By rescaling the atomic energy in such a way that $E_{g}=0$, with $E_{g}$ being the energy of the ground state of each atom, it is straightforward to see that the Hamiltonian of the system can be written so that the corresponding time-evolution operator reads $\hat{U}_{r}=e^{-2i\frac{\lambda^2}{\delta}t\hat{n}\pro{e}{e}}$. This form of the coupling Hamiltonian, which has been discussed in many contexts of quantum optics and quantum information processing~\cite{englert,ioMyungKnight}, is here reinterpreted as the generalization of the Ising interaction between two qubits. This latter has the structure $\hat{U}_{I}=e^{-i\chi{\hat{\sigma}}^{1}_{z}\hat{\sigma}^{2}_{z}}$, where $1$ and $2$ are the labels for the two qubits, $\chi$ is the interaction rate and $\sigma^{j}_{z}$ is the $z$-Pauli operator for qubit $j=1,2$. Evidently, neither $\hat{U}_{I}$ nor $\hat{U}_{r}$ describe an effective exchange of excitations between the subsystems they involve. Just a conditional phase-shift is applied, as a result of these two time-propagators, on the joint state of the two subsystems. As the Ising interaction is the basis of the theoretical approach to quantum information processing with cluster states~\cite{briegel}, this formal reinterpretation paves the way toward an interesting analysis of the quantum state transfer, quantum entanglement transfer and entanglement reciprocation in terms analogous to the mechanisms which rule the propagation and manipulation of information in a cluster state-based scenario. The following discussion, which is taken from the general cluster state formalism~\cite{briegel,noiCLUSTER}, serves as a building block for the description of the information entanglement transfer protocols with Eq.~(\ref{evol-op-2}).

In order to transfer a (normalized) single-qubit quantum state, $|\psi\rangle = a|0\rangle_{1} + b |1\rangle_{1}$ of qubit $1$ to qubit $2$ (logical states $\{\ket{0}\equiv\ket{g},\ket{1}\equiv\ket{e}\}_{j},\,j=1,2$ and $a,\,b\in\mathbb{R}$, for simplicity), we first prepare $2$ in the balanced superposition $|+\rangle=(|0\rangle+|1\rangle)/\sqrt{2}$. Then, we apply $\hat{U}_{I}$, which is equivalent to a control-phase operation that, for $\chi=\pi$, gives a phase-flip only when both the qubits are in $|1\rangle$. The joint system is thus in the quantum state $$|\Psi\rangle_{12} = a |0,+\rangle_{12} + b |1,-\rangle_{12}.$$

By performing a projective measurement on $\{|\pm\rangle_{1}\}$ (with $\ket{-}_{1}=(1/\sqrt 2)(\ket{0}-\ket{1})$), the output state of $2$ becomes $a|+\rangle_2 \pm b|-\rangle_2$, corresponding to the measurement outcome $\pm$ respectively. Depending on the outcome of the measurement, a conditional unitary operation on $2$ is applied, in order to retrieve perfectly the information initially encoded in the state of qubit $1$. For the outcome $+$, $\one$ has to be applied to qubit $2$ while, for the outcome $-$, the application of $\hat{\sigma}_x$ is required (where $\hat\sigma_x$ is the $x$-Pauli operator). After these conditional correction operations, the state of $2$ becomes $a|+\rangle+b|-\rangle$. Notice that the initial qubit is now encoded in the eigenstates $\ket{\pm}$ of $\hat{\sigma}_{x}$ rather than in the energy eigenstates. The conversion into this basis can be accomplished with the application of a {\it byproduct operator}~\cite{briegel,noiCLUSTER} corresponding to a Hadamard transformation, which rotates the state into the original basis $\{|0\rangle, |1\rangle\}$.

In complete analogy to the above discussion, a complete state transfer to a coherent state of the cavity field can be accomplished, through $\hat{U}_{r}$ in Eq.~(\ref{evol-op-2}), by applying a modified version of the above protocol. The initial information-encoded qubit $a\ket{0}+b\ket{1}$ is now allowed to interact with a coherent state of the cavity field $\ket{\alpha}$ through Eq.~(\ref{evol-op-2}) for an interaction time $\lambda^2{t}/\delta=\pi/2$, leading to the joint state

\begin{equation}
\label{statetransfer}
(a\ket{0}+b\ket{1})\ket{\alpha}\stackrel{\hat{U}_{r}}{\longrightarrow}\frac{1}{\sqrt{2}}(a\ket{g}\ket{i\alpha}-ib\ket{e}\ket{-i\alpha}).
\end{equation}

A projection of the qubit state onto the $y$-Pauli matrix $\hat{\sigma_{y}}$ eigenstates $\ket{\pm}_{y}=(1/\sqrt 2)(\ket{g}\pm{i}\ket{e})$ leads to the state ${\cal N}_{+}(a\ket{i\alpha}-b\ket{-i\alpha})$ (for an outcome $\ket{+}_{y}$) or ${\cal N}_{-}(a\ket{i\alpha}+b\ket{-i\alpha})$ (for an outcome $\ket{-}_{y}$), where ${\cal N}_{\pm}=(1\pm{2}abe^{-2|\alpha|^2})^{-1/2}$. By remarking that, already for $|\alpha|\ge{2}$, $\langle{i}\alpha|-i\alpha\rangle\simeq{10^{-4}}$, the resulting states represent information-encoded quasi-qubit states. Differently from the case where the information-receiver was a qubit, in this instance the wrong relative phase appearing in correspondence to the $\ket{+}_{y}$ outcome can be corrected by applying a ${z}$-rotation in the Hilbert space of the qubit spanned by $\{\ket{i\alpha},\ket{-i\alpha}\}$. This rotation can be formally performed by exploiting the technique developed in~\cite{jacob,ioMyungKnight}. In our cavity-QED setup, we need the injection of an external, amplitude-controlled coherent state into the cavity, which can be done as explained in Ref.~\cite{brune}. However, as our tasks are not related to quantum computation but to communication and information transfer, it is enough to assume here a conservative approach based on just the postselection of the events leading to the $\ket{-}_{y}$ outcomes.


\section{Entanglement transfer from atoms to CV}

\label{transfer}

The basic principles described in the previous Section can be adapted and exploited in order to transfer quantum correlations initially in a two-qubit state to the separable state of two cavity field-modes. The task is relevant and interesting as the creation of an entangled channel of a CV state is one of the focus points in quantum state engineering with CV. Recently, a considerable interest has been directed toward the achievement of particular regimes of interaction between matter and light to be used for the purposes of generating entangled coherent states~\cite{noiEIT}. On the other hand, here the approach is different as our task would be the {\it transfer} of correlations already present in an ancillary state onto the joint state of a bipartite mode-field. Such a scenario is made possible by the ease with which entangled pairs of atoms can be created in microwave cavity-QED, as reported in~\cite{Haroche97}, where the entangling technique has been also extended to the generation of three-particle entanglement.

\begin{figure}[t]
\psfig{figure=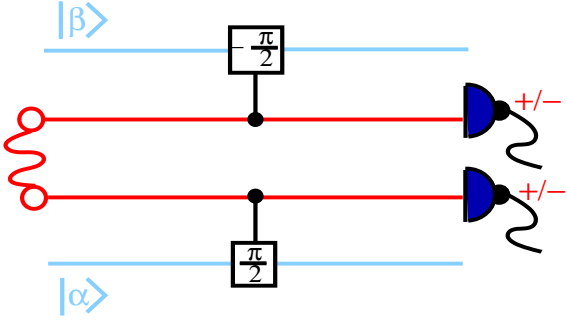,width=7.5cm,height=4.3cm}
\caption{Quantum circuit scheme of the protocol for entanglement transfer
from a pair of entangled qubits to two initially separable coherent states.
The wave-line represents the entanglement initially between the qubits (inner lines). The symbols for the conditional phase-shift gates implemented by the cross-Kerr nonlinear interactions are shown with the corresponding phase-shift. We also include the atomic-state detectors (which perform projections onto the eigenstates of $\hat{\sigma}_{x}$).}
\label{sketch1}
\end{figure}

We consider two atoms, labelled as $1$ and $2$, in the maximally entangled state (generated as an off-line resource)
\begin{eqnarray}
\label{eq:tqmes}
|\psi_+\rangle_{12} = \frac{1}{\sqrt{2}} \left(|eg \rangle +
|ge \rangle \right)_{12}.
\end{eqnarray}
As a {\it receiving} two-mode CV system we can consider either two separate single-mode microwave cavities or a single bimodal cavity. In the first situation, the protocol we are going to describe works with two cross-Kerr interactions performed in parallel. The second configuration, on the other hand, would imply a sequence of atomic passages. The atomic detunings can be set in such a way that, if we label the field-modes as $a$ and $b$, the effective interaction configuration would involve the pairs $(1,a)$ and $(2,b)$, with the interaction being ruled by Eq.~(\ref{evol-op-2}) for each pair. The initial state of the cavity fields is taken to be $ |\psi\rangle_{ab} = |\alpha, \beta\rangle_{ab}$ so that the composite system is in the state
\begin{eqnarray}
\label{eq:iscs}
|\Psi\rangle_{1a2b} = \frac{1}{\sqrt{2}} \left(|e, \alpha, g, \beta \rangle + |g, \alpha, e, \beta \rangle \right)_{1a2b}.
\end{eqnarray}
In what follows, $\delta_{Af}$ represents the atom-field detuning with $A=1,2$ and $f=a,b$. The working conditions we adopt here are such that, for a sequential passage of atoms $1$ and $2$ through a bimodal cavity, $|\delta_{1b}|,|\delta_{2a}|\gg|\delta_{2b}|=|\delta_{1a}|$. Obviously, this condition can be relaxed for a parallel realization involving two separate cavities. The atom-field interaction is arranged in order to satisfy the evolution $\hat{C}^{1a}_p\equiv\hat{U}_r(\frac{\lambda^2 t}{\delta_{1a}}=\frac{\pi}{2})$ and $\hat{C}^{2b\,\dagger}_p=\hat{U}_r(\frac{\lambda^2 t}{\delta_{2b}}=-\frac{\pi}{2})$, where the latter condition is possible by setting the detuning $\delta_{2b}<0$. The state of the composite system is thus given by
\begin{eqnarray}
\label{eq:acpafs}
\left |\Psi\left(\frac{\pi}{2}\right)\right\rangle = \frac{i}{\sqrt{2}} \left(-|e, -i \alpha, g, -i \beta\rangle + |g, i \alpha, e, i \beta \rangle \right) .
\end{eqnarray}
The information about the atomic state is now written onto the CV state. In order to {\it extract} just the state of the field-mode, we perform a measurement over the atomic part of $\ket{\Psi(\pi/2)}$ by projecting it on the basis $|\pm\rangle_{1}\ket{\pm}_{2}$ with $\ket{\pm}_{j} = (|e\rangle \pm |g\rangle)_{j}/\sqrt{2},\,(j=1,2)$. Ignoring the global phase of $i$, the output state of the two-mode field is given by
\begin{eqnarray}
{\cal N}_{\alpha,\beta,-}\left(|-i \alpha, -i \beta
\rangle - |i \alpha, i \beta \rangle \right)~~~&&\mbox{for }(\pm,\pm)
\label{eq:osotf} \\
{\cal N}_{\alpha,\beta,+}\left(|-i \alpha, -i \beta
\rangle +|i \alpha, i \beta \rangle \right)~~~&&\mbox{for }(\pm,\mp).
\label{17}
\end{eqnarray}
with ${\cal N}_{\alpha,\beta,\pm}$ being proper normalization factors. For a large enough amplitude of the coherent states involved in Eqs.~(\ref{eq:osotf}) and (\ref{17}), these states manifestly show that entanglement is present. However, we wish to know whether or not this entanglement is maximal, and so to establish the conditions for optimality of the entanglement transfer protocol, given that we started with an initial atomic ebit. In the following analysis we thoroughly investigate the conditions under which an ebit is encoded in a two-mode entangled coherent state.


\subsection{Complete ebit for an entangled coherent state}

\label{conditions}

For simplicity of the argument, let us assume two modes having the same amplitude $\gamma\equiv{i}\alpha=i\beta$ with $\gamma$ real. The states (\ref{eq:osotf}) and (\ref{17}) are then simply
\begin{eqnarray}
&&\frac{1}{\sqrt{2(1-\mbox{e}^{-4\gamma^2})}}(|-\gamma\rangle|-\gamma\rangle-|\gamma\rangle|\gamma\rangle)
\label{cat1}\\
&&\frac{1}{\sqrt{2(1+\mbox{e}^{-4\gamma^2})}}(|-\gamma\rangle|-\gamma\rangle+|\gamma\rangle|\gamma\rangle).
\label{cat2}
\end{eqnarray}
We find that the probability of getting the $(+,+)$ or $(-,-)$ outcome is $P_{++}=P_{--}=\frac{1}{4}(1-\mbox{e}^{-4\gamma^2})$ and that of getting the $(+,-)$ or $(-,+)$ outcome is $P_{+-}=P_{-+}=\frac{1}{4}(1+\mbox{e}^{-4\gamma^2})$. Let us consider two orthogonal bases  $|\phi\rangle={\cal N}_\phi(|\gamma\rangle+|-\gamma\rangle)$ and $|\psi\rangle={\cal N}_\psi(|\gamma\rangle-|-\gamma\rangle)$. It is straightforward to see that, for the measurement events $(\pm,\pm)$, the field modes (\ref{cat1}) can be written in the form
\begin{equation}
\frac{1}{\sqrt{2}}(|\phi\rangle|\psi\rangle+|\psi\rangle|\phi\rangle),
\end{equation}
which is a maximally entangled qubit state. We can get this maximally entangled state with probability $2P_{++}$.   However, it is not possible to find an effective orthonormal qubit basis such that the state (\ref{cat2}) is written as a maximally entangled state. Indeed, by using the same orthonormal basis exploited above, we get that Eq.~(\ref{cat2}) can be reduced to the form 
\begin{equation}
\frac{1}{\sqrt{{\cal N}^4_{\phi}+{\cal N}^4_{\psi}}}({\cal N}^2_{\psi}\ket{\phi}\ket{\phi}+{\cal N}^{2}_{\phi}\ket{\psi}\ket{\psi})
\end{equation}
with
${\cal N}_{\phi}=[2(1+e^{-2\gamma^2})]^{-\frac{1}{2}}$ and ${\cal N}_{\psi}=[2(1-e^{-2\gamma^2})]^{-\frac{1}{2}}.$
The von Neumann entropy (VNE) can be used as a measure of the entanglement in this pure effective two-qubit state. It is straightforward to see that VNE approaches unity for a large enough amplitude of the coherent state component of Eq.~(\ref{cat2}) whereas (\ref{cat1}) carries one ebit regardless of $\alpha$.    


More generally, the condition to assign a complete ebit for an entangled coherent state can be written as follows. A coherent superposition of a field mode $f=a,b$ is 
\begin{eqnarray}
\label{eq:ecs}
|\psi\rangle_f = a |\alpha\rangle_{f} + b |\beta\rangle_{f}.
\end{eqnarray}
The state may be described in a two-dimensional Hilbert space, spanned
by $\{|\alpha\rangle_f$, $|\beta\rangle_f\}$. The coherent states are
mutually nonorthogonal and one can find an orthonormal basis~\cite{Jeong01} such as
\begin{equation}
\label{eq:onb}
\begin{aligned}
|\psi_+\rangle_f=& \frac{1}{\sqrt{N_f}} \left(e^{-i\phi_f/2}\cos\theta_f
|\alpha\rangle_f - e^{i\phi_f/2} \sin\theta_f |\beta\rangle_{f} \right),\\
|\psi_-\rangle_f=& \frac{1}{\sqrt{N_f}} \left(-e^{-i\phi_f/2}\sin\theta_f
|\alpha\rangle_f + e^{i\phi_f/2} \cos\theta_f |\beta\rangle_f \right),\\
\end{aligned}
\end{equation}
where $N_f=\cos(2\theta_f)$, $\sin(2\theta_f)= |_f\langle \alpha|\beta\rangle_{f}|$ and $\exp[-i\phi_f]=_{f}\!\!\langle \alpha |\beta\rangle_f/|_f\langle \alpha |\beta \rangle_f|$. By inverting the above relations, the coherent states can be written as
\begin{equation}
\begin{aligned}
\label{eq:csrbonb}
|\alpha\rangle_f &=e^{i\phi_f/2}\cos\theta_f
|\psi_+\rangle + e^{i\phi_f/2} \sin\theta_f |\psi_-\rangle,\\
|\beta\rangle_f &=e^{-i\phi_f/2}\sin\theta_f
|\psi_+\rangle + e^{-i\phi_f/2} \cos\theta_f |\psi_-\rangle.
\end{aligned}
\end{equation}

Suppose that we have an entangled coherent state,
\begin{equation}
|\Phi\rangle \equiv |\alpha,\alpha\rangle_{ab} + e^{i \psi}
|\beta,\beta\rangle_{ab},
\end{equation}
neglecting the normalization.  This {\em pure} state carries one
ebit of entanglement if $\mbox{Tr}_b(|\Phi\rangle \langle\Phi|)\propto\openone$. Here $\mbox{Tr}_f$ stands for the partial trace over the two-dimensional Hilbert space of the mode $f$. This condition is equivalent to
\begin{equation}
\label{eq:mecfecs}
\begin{aligned}
\sin 2\theta_a&=\sin 2\theta_b,\\
\psi-\phi_a-\phi_b&=(2n+1)\pi,
\end{aligned}
\end{equation}
where $n\in\mathbb{Z}$. For a state like $\ket{\bar{\Phi}}=|-i \alpha, -i\alpha \rangle_{ab} + e^{\psi}|i \alpha, i \alpha \rangle_{ab}$, we have $\theta_a=\theta_b$ and $\phi_a=\phi_b=0$. Therefore, the condition that the entangled coherent state has one ebit is reduced to $\psi =(2n+1)\pi$.
More generally, a state having the form $|\Phi'\rangle=|\alpha, \beta \rangle_{ab} + e^{i\psi}|\beta, \alpha\rangle_{ab}$ caries one ebit if $\psi = 2\phi+(2n+1)\pi$, where $e^{-i\phi} =\langle \alpha |\beta \rangle/|\langle \alpha |\beta \rangle|$.

The above analysis shows that the postselection of the atomic detection events $(\pm,\pm)$ allows us to achieve the optimal performance of the entanglement transfer protocol. The initial atomic ebit is converted, by a process involving cross-Kerr-based QND interactions and orthogonal projection of qubit states, to an entangled coherent state carrying exactly one ebit.


\section{Entanglement reciprocation from CV to atoms}

\label{reciprocation}

In the previous Section, we have described a protocol to transfer an ebit of entanglement from a discrete variable to the joint state of two CV subsystems. The opposite process, which should allow the retrieval of the qubit entanglement, is denominated {\it entanglement reciprocation}~\cite{noireciprocation}. In the context of Ref.~\cite{noireciprocation}, it has been shown that a full ebit can be reciprocated to and from a CV state through the use of a resonant atom-field interaction and appropriate postselection processes. Here we extend these results to the case of the dispersive model we use and show that a full reciprocation mechanism is possible for a cross-Kerr interaction. On one hand, this analysis allows us to stress the flexibility of the cross-Kerr interactions in the context of quantum state engineering. On the other hand, the results of this investigation relieve the entanglement reciprocation mechanism from the specific interaction model considered in~\cite{noireciprocation}, showing its effectiveness for other, non-resonant interaction Hamiltonians. Once again, the basic building block will be the use of cross-Kerr interaction in effective information transfer processes.

In order to reciprocate the entanglement contained in a CV channel (such as an entangled coherent state) to two initially separable atoms, we consider a procedure entirely similar to the qubit$\rightarrow$CV entanglement transfer. We suppose the fields are in a state equivalent to Eq.~(\ref{eq:osotf}) which, as we have already remarked in Section~\ref{conditions}, contains exactly one ebit, and we prepare two atoms in the product state of two $\hat{\sigma}_{x}$ eigenstates
\begin{eqnarray}
\label{eq:taipsfcta}
|\psi\rangle_{12} = |+,+\rangle_{12}.
\end{eqnarray}
Now the state of the composite system is given by
\begin{eqnarray}
|\Psi\rangle_{1a2b}\propto \left(|+, -i\alpha, +, -i\beta \rangle - |+, i\alpha, +, i\beta \rangle \right)_{1a2b}.
\end{eqnarray}

\begin{figure}[t]
\psfig{figure=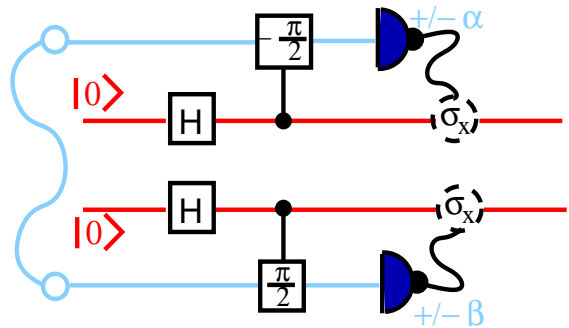,width=7.5cm,height=4.3cm}
\caption{Quantum circuit scheme of the protocol for entanglement reciprocation
between an entangled coherent state of two modes (outer lines connected by the wavy line) and two initially separable qubits. A complete ebit can be transferred from the field modes to the qubits. The symbols for the Hadamard gates and the conditional phase-shift gates implemented by the cross-Kerr nonlinear interactions are shown (these latter with the corresponding phase-shift). We also include the field-state detectors (which should discriminate between $\pm\xi$ with $\xi=\alpha,\beta$). Finally, some operations (in the form of $\hat{\sigma}_{x}$ rotations) are required, conditioned on the result of the field measurements and some classical communication.}
\label{sketch2}
\end{figure}

The next step is the application of an effective $\hat{C}_p^\dag$ gate to the pair $(1,a)$, while $\hat{C}_p$ is applied to the atom-field subsystem $(2,b)$. The state becomes
\begin{equation}
\label{eq:acpferfcta}
\begin{aligned}
|\Psi\rangle&\propto\left[\left(
-|e, e \rangle +|g, g \rangle \right) \left( |-\alpha,\beta\rangle -|\alpha,-\beta\rangle\right)\right.\\
  &\left.+i\left(|e, g \rangle +|g, e \rangle \right) \left( |\alpha,\beta\rangle -|-\alpha,-\beta\rangle\right) \right].
\end{aligned}
\end{equation}

In order to disentangle the CV state from the qubit state we perform a measurement of the field modes by projecting them onto $\{|\pm \alpha \rangle\}$ and $\{|\pm \beta\rangle\}$. If the phase of both the fields is the same, the atoms are projected onto the maximally entangled state Eq.~(\ref{eq:tqmes}).
On the other hand, if the phases are different, a conditional unitary operation $\hat{\sigma}_y$ on the second atom has to be applied in order to retrieve the form (\ref{eq:tqmes}). Here we assume a measurement that can perfectly discriminate the states $\{|\pm \alpha \rangle\}$ and $\{|\pm \beta \rangle\}$). While an effective procedure which can implement such a measurement is briefly described in the next Section, here we stress that the efficiency of this scheme obviously depends on the measurements performed on the field modes, whose success probability depends on the magnitudes of $\alpha$ and $\beta$. If $|\alpha|\gg 1$ and $ |\beta| \gg 1$, the success probability approaches unity. Correspondingly, the final atomic state will be pure and carry one ebit. Otherwise, if the success probability is smaller, the corresponding atomic state is mixed and only partially entangled. In order to quantify the efficiency of the protocol and its resilience against this sort of imperfection, we exemplify the results of a protocol where $\{\ket{\alpha},\ket{\beta}\}$ are measured. The other cases can be treated in a manner equivalent to the approach we are going to describe.

We model the measurement as a projection of the state~(\ref{eq:acpferfcta}) performed by the operator
\begin{equation}
\label{proiettore}
\hat{\Pi}_{a}\otimes\hat{\Pi}_{b}=\int{d}\gamma\int{d}\delta\ket{\gamma}_{a}\!\bra{\gamma}\otimes\ket{\delta}_{b}\!\bra{\delta}G_{\gamma}(\Delta,\alpha)G_{\delta}(\Delta,\beta),
\end{equation}
where $G_{\mu}(\Delta,\chi)$ ($\mu=\gamma,\delta$ and $\chi=\alpha,\beta)$ are proper (normalized) weighting function centered on $\chi$ and having width $\Delta$. This model accounts for the detection of a coherent state which is not exactly the state we would like to detect. For the sake of simplicity, we can assume each weigthing function to be a normalized Gaussian $G_{\mu}(\Delta,\chi)=\frac{1}{\sqrt{2\pi\Delta}}e^{-\frac{(\mu-\chi)^2}{2\Delta}}$. It is obvious, however, that the model~(\ref{proiettore}) is independent of the actual choice for these functions.

It is straightforward to prove that the normalized conditioned qubit density matrix, after the detection event, is given by
\begin{equation}
\label{dopomisura}
\begin{aligned}
\rho_{12}=&\frac{1}{P_1+P_2}[P_{1}\ket{\phi_-}_{12}\!\bra{\phi_{-}}+P_2\ket{\psi_{+}}_{12}\!\bra{\psi_+}\\
&-iP_{3}(\ket{\phi_-}_{12}\!\bra{\psi_+}-\ket{\psi_+}_{12}\!\bra{\phi_-})]
\end{aligned}
\end{equation}
where $\ket{\phi_-}_{12}=(1/\sqrt{2})(\ket{gg}-\ket{ee})_{12}$, $\ket{\psi_+}_{12}=(1/\sqrt{2})(\ket{ge}+\ket{eg})_{12}$ and $P_{j}\,(j=1,2,3)$ are real functions depending on the overlaps $\langle{\pm\alpha}\vert\gamma\rangle$ and $\langle\pm\beta\vert\delta\rangle$ (for definiteness we assume here $\alpha,\beta\in\mathbb{R}$). We are interested in the fidelity ${\cal F}_{\psi_{+}}\!=_{12}\!\bra{\psi_{+}}\rho_{12}\ket{\psi}_{12}$ which estimates how close the two-qubit state is, after the measurement to $\ket{\psi}_{12}$. Analytically, we find
\begin{equation}
\label{fide}
{\cal F}_{\psi_{+}}=\frac{1+e^{-\frac{4(\alpha^2+\beta^2)}{1+2\Delta}}-2e^{-\frac{2(1+\Delta)(\alpha^2+\beta^2)}{1+2\Delta}}}{(P_1+P_2)(1+2\Delta)}
\end{equation}
with
$$P_{1}=\frac{1}{1+2\Delta}({e^{-\frac{4\alpha^2}{1+2\Delta}}+e^{-\frac{4\beta^2}{1+2\Delta}}-2e^{-\frac{2(1+\Delta)(\alpha^2+\beta^2)}{1+2\Delta}}}),$$
$$P_{2}=\frac{1}{1+2\Delta}({1+e^{-\frac{4(\alpha^2+\beta^2)}{1+2\Delta}}-2e^{-\frac{2(1+\Delta)(\alpha^2+\beta^2)}{1+2\Delta}}}).$$

\begin{figure}[ht]
\psfig{figure=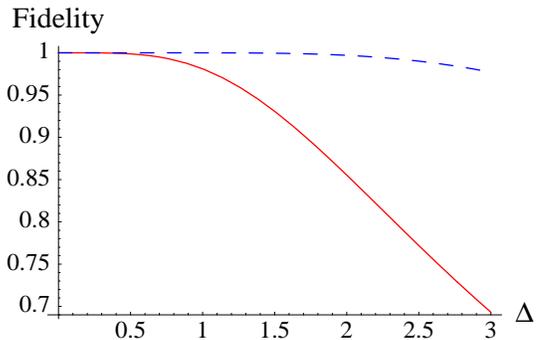,width=7.5cm,height=4.5cm}
\caption{Fidelity of the conditioned atomic state against the width $\Delta$ of the weigthing functions in Eq.~(\ref{proiettore}). Here, we show the results for $\alpha=\beta=2$ (solid line) and $\alpha=\beta=3$ (dashed line). The robustness of the scheme for reciprocation is dramatically improved for properly large amplitudes of the coherent states.}
\label{fede}
\end{figure}

The behavior of the fidelity against the width $\Delta$, for two choices of the amplitudes $\alpha,\beta$ of the coherent states, is shown in Fig.~\ref{fede}.
A considerable robustness of the scheme against imperfections in the CV-state detection events is made evident for a proper choice of the amplitude of the components of the CV part. Plots showing analogous trends can be obtained regarding the entanglement in the atomic state and its purity: for $\alpha,\beta\ge{3}$ a highly pure state carrying nearly one ebit of entanglement can be obtained for values of $\Delta$ up to $3$. The corresponding state is extremely close to $\ket{\psi_+}_{12}$, as shown by a fidelity-based analysis.


\section{Entanglement transfer to CV from Multiple Entangled Atomic Pairs and Reciprocation}
\label{multiple}

The entanglement transfer and reciprocation schemes discussed in
this paper can be easily extended to a multiple set of entangled
atomic pairs. The trick is simply to make successive pairs of
entangled atoms interact with their respective cavity fields for
distinct amounts of time. We start by representing the joint
entangled state of $n$ atoms as
\begin{eqnarray}
\label{eq:many}
|\Psi\rangle_{12} =|\psi_{+}\rangle_{12}^{\otimes n}=\frac{1}{2^{n/2}}\sum_{x=0}^{2^n-1}|x,x^c\rangle_{12},
\end{eqnarray}
where we have used the labeling
$|e\rangle=|0\rangle,|g\rangle=|1\rangle$, and $x=x_1x_2...x_n$ and
$x^c=x_1^cx_2^c...x_n^c$ are binary numbers. If we start with the
cavity fields in $|\alpha\rangle_a$ and $|\alpha\rangle_b$, and make
each pair of atoms interact with the cavity fields for half the time
as the previous one in succession, we can obtain the state
\begin{eqnarray}
\label{eq:many1}
|\Psi(t)\rangle_{12,ab} =\sum_{x=0}^{2^n-1}|x,x^c\rangle_{12}|\alpha(x)\rangle_a|\alpha(x^c)\rangle_{b},
\end{eqnarray}
where $\alpha(x)=\alpha e^{i\pi(x_1/2+x_2/4+...+x_n/2^n)}$. Next all
the atoms are measured in the $|\pm\rangle$ basis. Say the outcome
is $|+\rangle^{\otimes n}_1|+\rangle^{\otimes n}_2$. Then the
cavities are projected to the state
\begin{eqnarray}
\label{eq:many2}
|\Psi(t)\rangle_{ab}=\sum_{x=0}^{2^n-1}|\alpha(x)\rangle_a|\alpha(x^c)\rangle_{b}.
\end{eqnarray}

If $\alpha\gg{1/}\sin(\pi/2^n)$, then the above is a $2^n\times
2^n$ dimensional maximally entangled state. In other words all the
entanglement of the initial $n$-ebits stored in the atomic states
have been transferred to the fields. All the other outcomes also
give the same type of entangled state apart from different relative
signs between the superposed components in $|\Psi(t)\rangle_{ab}$.
For reciprocation of the entanglement back to the atoms, a set of
$n$ atoms in each cavity are first prepared in the state $|+\rangle^{\otimes n}_1|+\rangle^{\otimes n}_2\equiv(\sum_{x=0}^{2^n-1}|x\rangle_1)(\sum_{x=0}^{2^n-1}|x\rangle_2)$. The previous pattern of interactions is now repeated, but with opposite sign of the interaction (this can be achieved by a $|e\rangle \leftrightarrow |g\rangle$ followed by the original interaction). Only for the term $\sum_{x=0}^{2^n-1}|x,x^c\rangle_{12}$ in the expansion of ($\sum_{x=0}^{2^n-1}|x\rangle_1)(\sum_{x=0}^{2^n-1}|x\rangle_2)$, both cavity fields are going to return to their original state $|\alpha\rangle_a\otimes|\alpha\rangle_b$. Thus by measuring the cavity fields, and obtaining the outcome $|\alpha\rangle_a\otimes|\alpha\rangle_b$ (again assuming $\alpha\gg{1}/\sin(\pi/2^n)$, so that this measurement is nearly one to an orthogonal basis), one projects the atoms to the $n$ pairs of ebits entangled state $|\Psi\rangle_{12}$. Thus the full amount of entanglement can be reciprocated conditional on an appropriate outcome.


\section{Entanglement swapping}

\label{swapping}

Let us suppose that a two-mode entangled coherent state of the form ${\cal N}_{\alpha,\alpha,-}(\ket{\alpha}\ket{\alpha}-\ket{-\alpha}\ket{-\alpha})_{a,b}$ has been created (for instance through the scheme described in Section~\ref{transfer}) between modes $a$ and $b$. On the other hand, we assume we have at our disposal a maximally entangled qubit state, of the form in Eq.~(\ref{eq:tqmes}), of qubits $1$ and $2$. The joint state will read
\begin{equation}
\label{swap}
\begin{aligned}
\frac{{\cal N}_{\alpha,\alpha,-}}{\sqrt{2}}&(\ket{e,\alpha,g,\alpha}-\ket{e,-\alpha,g,-\alpha}\\
&+\ket{g,\alpha,e,\alpha}-\ket{g,-\alpha,e,-\alpha})_{1,a,2,b}.
\end{aligned}
\end{equation}
We arrange a $\hat{C}_{p}\equiv\hat{U}_{r}(\pi/2)$ interaction between qubit $2$ and mode $b$ which will change the state above into
\begin{equation}
\label{swap2}
\begin{aligned}
\frac{{\cal N}_{\alpha,\alpha,-}}{\sqrt{2}}&(\ket{e,\alpha,g,i\alpha}-\ket{e,-\alpha,g,-i\alpha}\\
&-i\ket{g,\alpha,e,-i\alpha}+i\ket{g,-\alpha,e,i\alpha})_{1,a,2,b}.
\end{aligned}
\end{equation}
\begin{figure}[t]
\psfig{figure=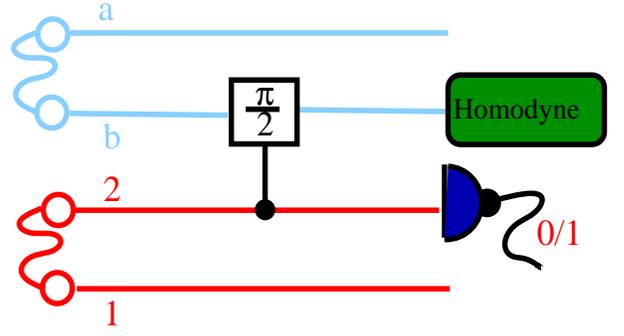,width=8.0cm,height=4.3cm}
\caption{Quantum circuit scheme of the protocol for entanglement swapping. The wavy lines represent the entanglement initially present in the bipartite state of modes $a$ and $b$ and between the qubits $1$ and $2$. We also show the symbol for the conditional phase gate between atom $2$ and mode $b$. The swapping process is completed by the measurement of these two subsystems. The measurement of mode $b$, in particular, is performed by a homodyne detection scheme as described in the body of the paper.}
\label{sketch3}
\end{figure}
A homodyne detection discriminating between the different possible phases of mode $b$ will project Eq.~(\ref{swap2}) onto either $(\ket{e,\alpha,g}+i\ket{g,-\alpha,e})_{1,a,2}$ or $(\ket{e,-\alpha,g}-i\ket{g,\alpha,e})_{1,a,2}$. Then, a measurement of qubit $2$ on the eigenbasis of $\hat{\sigma}^{2}_{y}$ projects these states onto superpositions which can be always reconducted (through local atomic unitaries) to the prototypical form
\begin{equation}
\label{swap3}
\frac{1}{\sqrt 2}(\ket{g,\alpha}+\ket{e,-\alpha}).
\end{equation}
The state in Eq.~(\ref{swap3}) falls into the category of states which can be represented by the general relation $\ket{g,\alpha}+\ket{e,e^{i\varphi}\alpha}$ with $\varphi\in[0,\pi]$. The entanglement in this class of states has been studied in detail in Ref.~\cite{ioJacobMyung}. It is possible to show that, for $\varphi=\pi$ and $|\alpha|\ge{2}$, the von Neumann entropy corresponding to the state (\ref{swap3}) is exactly one ebit. This implies that, under these conditions, Eq.~(\ref{swap3}) represents a maximally entangled state, therefore implying that the entanglement swapping procedure implemented by the scheme presented here is complete.

The homodyning of the cavity field-mode $b$ can be performed by injecting an external coherent state of amplitude $i\alpha$ into the cavity. This will add to the pre-existent coherent state in such a way that either the state $\ket{0}_{b}$ or $\ket{2i\alpha}_{b}$ is obtained. Then a resonant interaction lasting for a rescaled interaction time $\pi$ with an ancillary atom prepared in its ground state $\ket{g}$ is required. If the field-mode was in its vacuum state, the probability of finding the atom in $\ket{g}$ will be strictly $1$. On the other hand, for a large enough $|\alpha|$, such a probability is smaller (being still non-zero because of the overlap between $\ket{0}$ and a coherent state), allowing for the discrimination of the phase of the cavity field.

The realization of the hybrid quantum channel in Eq.~(\ref{swap3}) can find application in the realization of quantum repeaters~\cite{repeaters} between cavity modes. Suppose, indeed, that the atomic flying qubit interacts with another cavity field-mode $c$ of an independent cavity. The task is the realization of an entangled channel of modes $a$ and $c$ which allows for long-haul quantum communication. The state transfer procedure described in Section~\ref{state} can be used in order to achieve such a result and create a two-mode coherent entangled state of $a$ and $c$. The realization of this scheme requires the passage of a single atom through multiple cavities, which is foreseeable with present state of the art microwave cavity-QED technology~\cite{raimond}.


\section{Conclusions}
\label{conlcusioni}

We have addressed the problem of quantum state engineering with cross-Kerr nonlinearities implemented in a cavity-QED based setup. This has allowed us to address not only the case of systems spanning Hilbert spaces of the same dimension but also the interesting situation of a hybrid system of discrete and CV variables. In this context, we have addressed different problems, such as entanglement transfer and reciprocation in order to build up discrete and CV channels for quantum communication and computation. Entanglement swapping between a discrete and a CV maximally entangled state has been addressed in order to consider the possibility of constructing maximally entangled hybrid channels that can have useful application in the context of quantum repeaters.

\acknowledgments

We thank the UK EPSRC, KRF (2003-070-C00024), The Leverhulme Trust
(ECF/40157) and the Korean Ministry of Science and Technology through
Quantum Photonic Science Research Center.

\end{document}